\documentclass[runningheads]{llncs}
\usepackage[T1]{fontenc}
\usepackage{graphicx}
\usepackage{amssymb}
\usepackage{amsmath}
\usepackage{multirow}
\usepackage{bbm}
\usepackage{gensymb}
\usepackage{booktabs}
\usepackage{subcaption}
\usepackage[section]{placeins}
\usepackage{tablefootnote}
\usepackage{listings}
\usepackage{placeins}
\usepackage{float}
\usepackage{tabularx}
\usepackage{tikz}

\usepackage{xcolor}
\usepackage{etoolbox,siunitx}
\robustify\bfseries
\usepackage[sort,nocompress]{cite}

\begin{document}
\title{Context-Aware Transformers For Spinal Cancer Detection and Radiological Grading}
\titlerunning{Spinal Context Transformer}
\author{Rhydian Windsor\inst{1} \and
Amir Jamaludin\inst{1}\and
Timor Kadir\inst{1,2}\and
Andrew Zisserman\inst{1}}
\authorrunning{Windsor et. al.}
\institute{Visual Geometry Group, Department of Engineering Science, University of Oxford \and
Plexalis Ltd, Oxford\\
\email{rhydian@robots.ox.ac.uk}\\
}
\maketitle      

\begin{abstract}
This paper proposes a novel transformer-based model architecture 
for medical imaging problems involving analysis of vertebrae.
It considers two applications of such models in MR images: 
(a) detection of spinal
metastases and the related conditions of vertebral fractures and  metastatic cord compression,
(b) radiological grading of common degenerative changes in intervertebral discs. 
Our contributions are as follows:
(i) We propose a Spinal Context Transformer (SCT), a deep-learning architecture 
suited for the analysis of repeated anatomical structures in medical imaging 
such as vertebral bodies (VBs). Unlike previous related methods, SCT considers all 
VBs as viewed in all available image modalities together, making predictions for 
each based on context from the rest of the spinal column and all available imaging modalities.
(ii) We apply the architecture to a novel and important task -- detecting spinal 
metastases and the related conditions of cord compression and 
vertebral fractures/collapse from multi-series spinal MR scans. 
This is done using annotations extracted from free-text radiological reports
as opposed to bespoke annotation. However, the resulting model 
shows strong agreement with vertebral-level
bespoke radiologist annotations on the test set.
(iii) We also apply SCT to an existing problem -- radiological grading of 
inter-vertebral discs (IVDs) in lumbar MR scans for common degenerative changes. 
We show that by considering the context of vertebral bodies in the image, 
SCT improves the accuracy for several gradings compared to previously published models.

\end{abstract}
\section{Introduction}
When a radiologist is reporting on a patient's imaging study, they will typically
make assessments based on multiple scans of the same patient. For example,
in a MRI study they will often consider multiple pulse sequences showing 
the same region and note differences between them. Related
conditions may appear similar in one modality and only be distinguishable when
more context is given via an additional imaging modality. Similarly, when 
reporting on spinal images, radiologists can learn a lot about a specific vertebra 
based on its appearance~\textit{relative to the vertebrae  neighbouring it}. For example,
a region of hyperintensity may represent a lesion or may just be an artifact of the scanning
protocol used; this often can be elucidated by looking at the rest of the spine.

Accordingly, automated models for spinal imaging tasks should also be able to make
vertebra-level predictions with context from multiple sequences \textit{and from neighbouring vertebrae}.
In this work, we propose the {\em Spinal Context Transformer} (SCT), a model
which aims to do exactly this. Crucially, SCT leverages 
light-weight transformer-based models which allow for variable numbers of 
inputs (both during training and at test-time) in terms of both:  (a) the number of 
vertebral levels, and (b) the number of input modalities depicting each vertebra. 

As well as introducing SCT, this paper also explores a new application 
of deep learning in medical imaging: automated diagnosis of spinal metastases 
and related conditions.
This is an important task for several reasons.
Firstly, metastases are very common; overall 2 in 3 cancer cases will metastasise~\cite{shaw_one-stage_1989}
(rising to nearly 100\% incidence in patients who die of cancer~\cite{maccauro_physiopathology_2011}). 
The spine is one of the most common places for these metastases to occur~\cite{ortiz_gomez_incidence_1995}.
Secondly, early detection is vital; metastases generally indicate advanced cancer
and, if left untreated, cause conditions such as vertebral fractures and 
metastatic cord compression, both of which can lead to significant pain and disability for 
the patient and are difficult and expensive to treat~\cite{van_tol_costs_2021,van_tol_time_2021}. 
Current clinical practice relies on heuristic-based scoring systems to 
quantify the development of metastases, such as SINS~\cite{fisher_novel_2010} and 
the Tokuhashi score~\cite{tokuhashi_scoring_2014}.
We aim to develop models to aid in the rapid detection and consistent quantification 
of this serious condition.
To further validate our model, we also test it on a previously published dataset for grading 
common intervertebral disc (IVD) degenerative changes in lumbar MR scans.

Another theme explored in this work is the ability of free-text radiological reports
to generate supervisory signals for medical imaging tasks. The common praxis
when training models on radiological images is to extract a relevant dataset from an imaging
centre, pseudo-anonymise it and then annotate the dataset for a condition of interest.
While this method is simple, effective and very popular, the additional step 
of annotating the dataset can be a significant bottleneck in 
the size of datasets used. This is a more serious problem than in conventional 
computer vision datasets as medical image annotation generally requires a specialist, 
whose time is limited and expensive. A much more scalable method of curating annotated datasets is to use
existing hospital records for generating supervisory signals. However, in radiology, these records are
usually in the form of free-text reports. Due to variation and lack of structure
in reporting, extracting useful information from these reports has been
considered a difficult task~\cite{casey_systematic_2021}, and, as such, most 
studies annotate their own data retrospectively. In this paper, we explore 
how useful the information in free-text reports 
can be for training models for our specific task of detecting spinal 
cancer,  and propose methods for dealing with ambiguities and omissions from the text.

\section{Related Work} 
The SCT, and its training and application, builds on three related areas of work: deep learning for analysis
of spinal scans, obtaining supervision from free-text radiology reports and transformers for aggregating
multiple sources of information. 
Automated detection and labelling of vertebrae is a long standing task in spinal MRI~\cite{Lootus13}
and CT~\cite{glocker_automatic_2012} analysis,
with deep learning now the standard approach~\cite{cai_multi-modality_2015,forsberg_detection_2017,tao_spine_2021, Windsor20b, Windsor20}.
One study particularly relevant to this work is that of Tao {\it et al.}~\cite{tao_spine_2021} 
who use a transformer architecture for this task. 
Beyond vertebra detection and labelling, deep learning has also been used to 
assess spinal MRIs for degenerative changes. 
For example, SpineNet of Jamaludin {\it et al.}~\cite{Jamaludin17b} and Windsor {\it et al.}~\cite{Windsor22} is a 
multi-task classification model acting on intervertebral discs which grades
for multiple common conditions. We compare to this approach here. 
DeepSpine of Lu {\it et al.}~\cite{lu_deepspine_18} and 
also Lewandrowski {\it et al.}~\cite{lewandrowski_feasibility_2020} 
use ground truth derived from radiological reports 
to train models for grading 
stenosis and disc herniation respectively, a theme we also explore in this work. In
terms of spinal cancer, \cite{zhao_discriminative_2020} and~\cite{wang_multi-resolution_2017}
both train models to detect metastases based on 2D images extracted from MRI studies;
and~\cite{merali_deep_2021} proposes a model to automatically detect 
metastatic cord compression, although this is restricted to axial T2 scans or the cervical
region. Several other works have considered vertebra and metastases detection
in other modalities such as CT (e.g.~\cite{hammon_automatic_2013,burns_automated_2013,glocker_automatic_2012,glocker_vertebrae_2013}),
although MR remains the clinical gold-standard in early spinal cancer detection. 
Finally, our approach also has parallels with works in video sequence analysis 
that combine a 2D CNN backbone with a temporal transformer
for tasks such as representation learning, tracking, object detection and 
segmentation~\cite{Bain21,gabeur2020mmt,carion_detr_2020,meinhardt2021trackformer}; though in our case,  visual features are extracted 
from multiple vertebrae shown in multiple MR sequences rather than consecutive frames.

\section{Spinal Context Transformer}
\begin{figure}[h!]
    \begin{subfigure}{\linewidth}
        \centering
        \includegraphics[width=.8\linewidth]{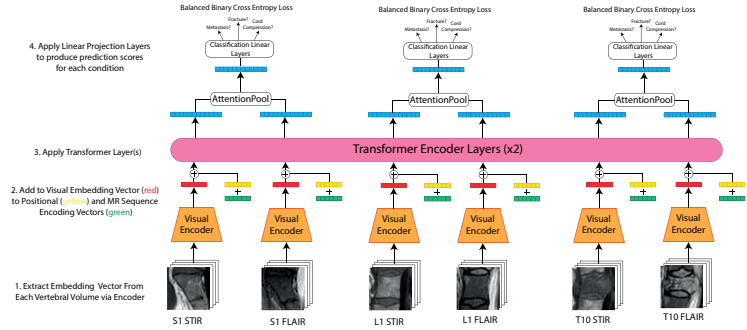}
        \caption{The SCT architecture
        illustrated for the spine cancer prediction task operating on multiple MRI sequences
        of three vertebrae. In the actual cancer detection task, the whole spinal column is used,
        and SCT is able to handle arbitrary variations in the number and type of input sequences and 
        vertebrae without adaptation. 
        }
        \label{fig:vert-transformer-full-architecture}
    \end{subfigure}
    \begin{subfigure}{\linewidth}
        \centering
        \includegraphics[width=0.7\linewidth]{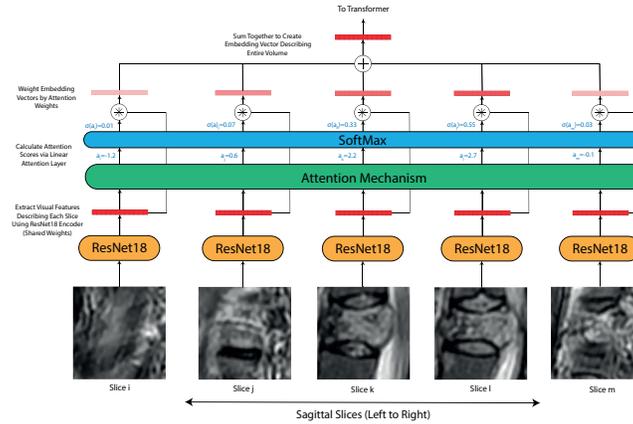}
        \caption{The visual encoder network used to extract features from each 
        vertebral volume, 
        This model can be trained by itself or 
        in conjunction with the transformer architecture as shown in (\subref{fig:vert-transformer-full-architecture}).
        The same model weights are used for all MR sequences.}

        \label{fig:vert-transformer-encoder}
    \end{subfigure}
    \caption{Architecture diagrams for SCT:  (\subref{fig:vert-transformer-full-architecture}) the full SCT architecture;
    (\subref{fig:vert-transformer-encoder}) the encoder
    used to extract visual features from each vertebra. STIR and FLAIR sequences are shown here, however the radiological
        grading task uses strictly T1 and T2 sequences, whereas the spinal cancer task uses a range of common sequences
        (T1, T2, STIR, FLAIR etc\.).
    }

\end{figure}

Like other models used to assess vertebral disorders such
as SpineNet \cite{Jamaludin17b, Windsor22} and DeepSpine~\cite{lu_deepspine_18}, the Spinal Context Transformer (SCT) 
exploits the repetitive nature of the spine by using the same CNN model to 
extract visual features from each vertebra given as input.
However, there are two key conceptual differences between how SCT and these other models 
operate which are described in this section.

Firstly, instead of using 3D convolutions and pooling layers to extract features from
multislice MR images, SCT instead extracts features from each slice
independently using a standard 2D ResNet18~\cite{He2016DeepRL} architecture adapted for single-channel images. 
The extracted feature
vectors for each slice are then aggregated using an attention mechanism. 
This is shown in Figure~\ref{fig:vert-transformer-encoder}. This design
choice is made for several reasons:  
(i) 3D convolutions assume all images have the same slice 
thickness. In real-world clinical practice a variety of scanning protocols are
used. This means slice thickness can vary significantly from sample to sample.
Of course, resampling can be employed in pre-processing to ensure a consistent
slice thickness for input images, however this introduces visual artifacts which may degrade
performance. 
(ii) Trained encoder models can operate on 
2D or 3D scans without adaptation. This also means image CNNs trained on large 2D datasets 
such as ImageNet~\cite{Deng09} can be used to initialize the encoder backbone.
(iii) The attention mechanism forces the model to explicitly
determine the importance of each slice for its prediction. This allows peripheral slices with partial-volume
effects to be ignored and can be useful
when it comes to interpreting the output of the model. An example of this effect
is shown in the appendix.
Secondly, as well as using attention to aggregate feature vectors from each slice,
SCT also uses attention to collate information from each MR sequence
and vertebra in the spinal column. This is done by feeding the vertebra's visual 
embedding vectors from each MR sequence to a lightweight 2-layer transformer 
encoding model, with additional embedding vectors describing the level of each input
vertebra and the imaging modality used. Both these additional embeddings
are calculated by linear layers operating on a one-hot encoding of the vertebra's name and sequence.
The output feature vectors for the same vertebra shown in different
sequences are then pooled together using a final attention mechanism identical to that 
used to pool features across slices. This creates a single output vector for
each vertebra. Linear classification layers convert this output vector into predictions 
for each  classification task. The full process is shown in Figure~\ref{fig:vert-transformer-full-architecture}.
Annotated PyTorch-style pseudo-code is given in the appendix.

\section{Detecting Spine Cancer Using Radiological Reports}
\label{sec:spinal-cancer}
This section describes applying SCT to a novel task:  detecting the presence of spinal
metastases in whole spine clinical MR scans,  as well as the related conditions
of vertebral fractures/collapse  and spinal cord compression, employing information from
free-text radiological reports for supervision. 
We use a dataset of anonymised clinical MRI scans and associated reports extracted from a local hospital 
trust PACS (Oxford University Hospitals Trust)
with appropriate ethical clearance.
Inclusion criteria were that the patient was over
18 years old, had at least one whole spine study,  and was referred from a cancer-related
specialty. In order to ensure a mix of positive and negative cases, all patients that
matched this criteria from April 2015 until April 2021 were included in the dataset, regardless of whether
or not they had metastases. 

To avoid having to annotate each scan independently we instead rely 
on the radiological reports written at the time the scan 
was taken to provide supervision for training our models.
A major advantage of this approach is that
it is much faster and thus readily scalable to larger datasets --  since no radiologist
is required to review each image independently,  an annotation set can be
obtained much quicker than would otherwise be possible. Furthermore, significantly
less clinical expertise is required to generate these annotations -- 
only a basic understanding of vocabulary related to the conditions of interest is needed.

\begin{figure}[h]
    \centering
    \begin{minipage}[h]{1.0\linewidth}
    \centering
        
    \begin{subfigure}{0.3\linewidth}
    \centering
    \includegraphics[width=\linewidth]{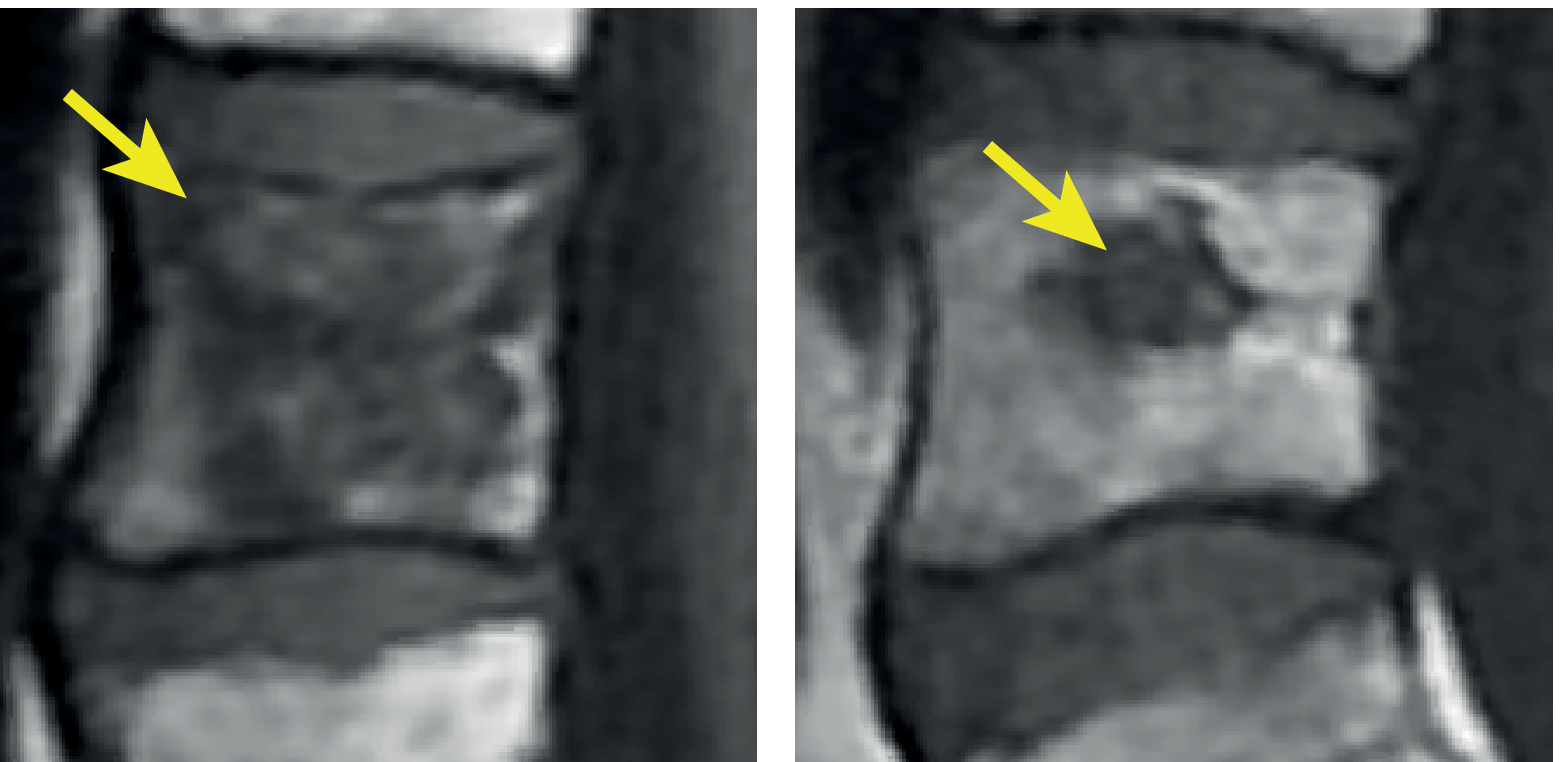}
    \caption{Metastases}        
    \end{subfigure}
    \begin{subfigure}{0.3\linewidth}
    \centering
    \includegraphics[width=\linewidth]{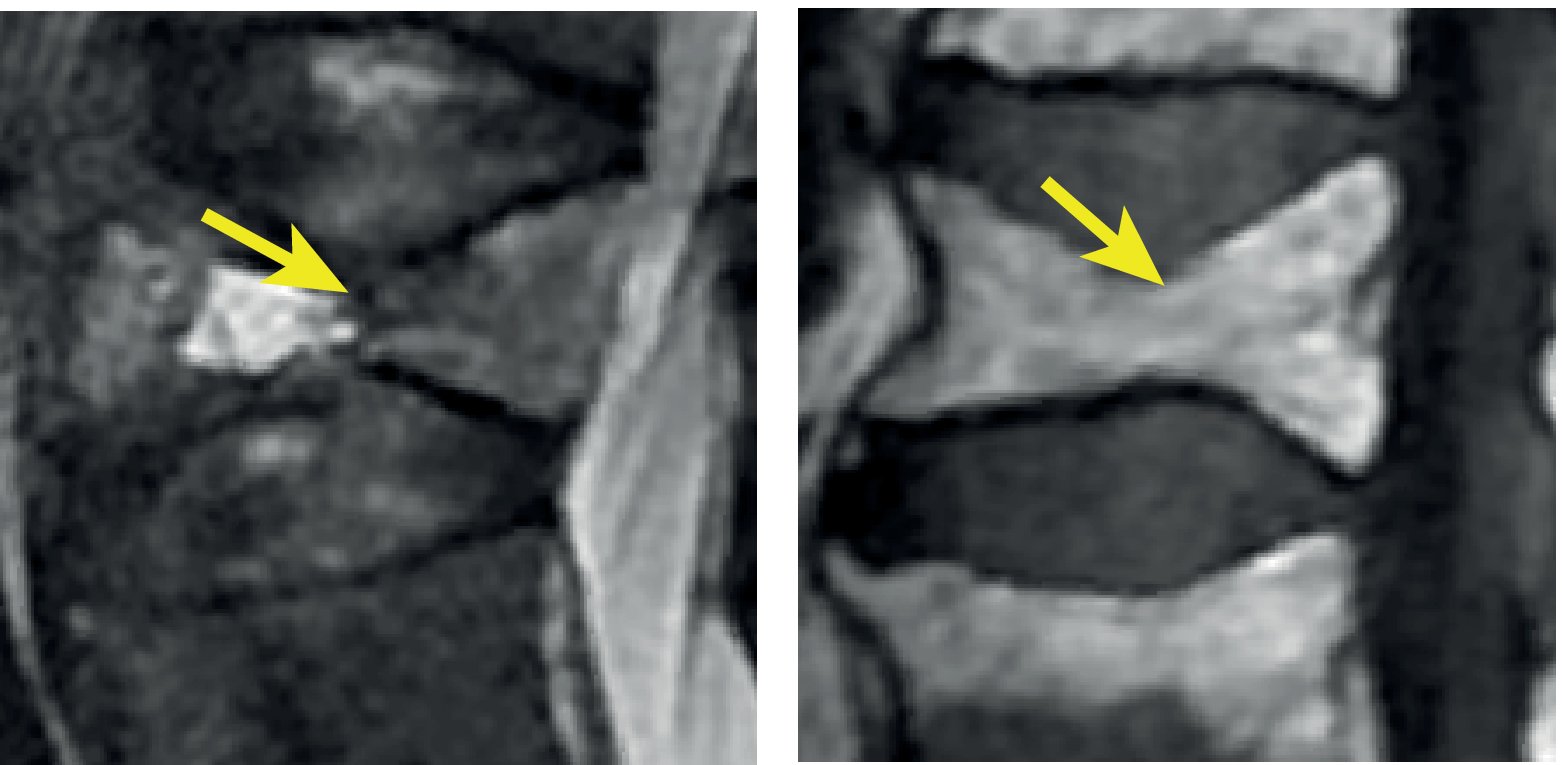}
    \caption{Collapsed Vertebrae}        
    \end{subfigure}
    \begin{subfigure}{0.3\linewidth}
    \centering
    \includegraphics[width=\linewidth]{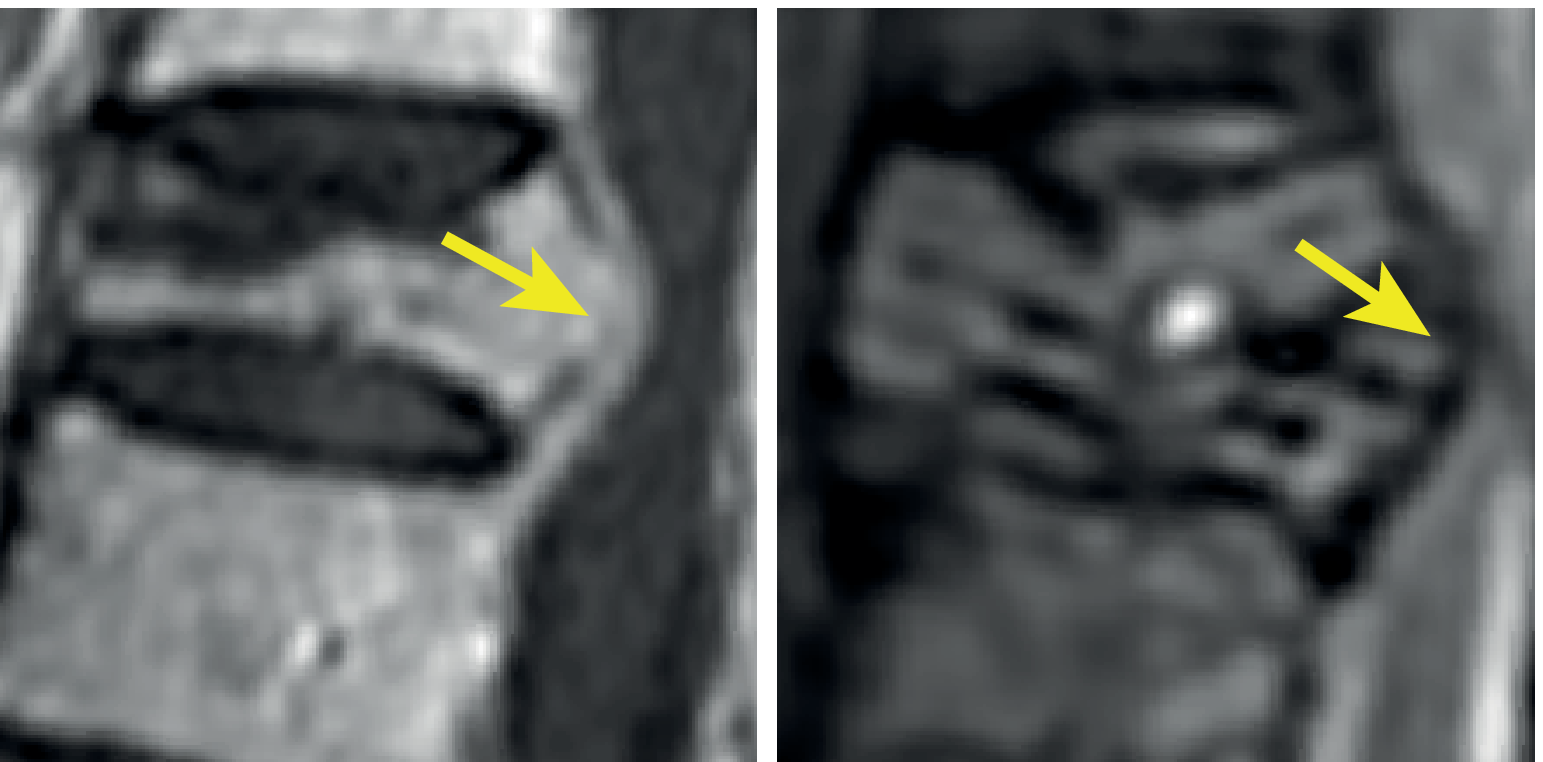}
    \caption{Cord Compression}        
    \end{subfigure}
    \caption{Example vertebrae with the three spinal-cancer related 
    conditions of interest. Note that in many cases vertebrae have a combination of
    these conditions.}
    \label{fig:example-cancer-annotations}

    \begin{footnotesize}
    \begin{tabular*}{\textwidth}{@{\extracolsep{\fill}}ccccccccccccc}
    \hline
    \multirow{3}{*}{Split} & \multirow{3}{*}{Patients} & \multirow{3}{*}{Studies} &  \multicolumn{9}{c}{Vertebral Bodies} \\
                                   &                           &                          & Total & \multicolumn{3}{c}{Metastases} & \multicolumn{3}{c}{Fractures} & \multicolumn{3}{c}{Compression}   \\
                                   &                           &                          &  & + &$-$ &? & + &$-$ &?& + &$-$ &?   \\
    \hline

    Training    & 258 & 558 & 12459 & 829 & 6099 & 5531 & 216&12124&119 & 90&12369&0 \\ 
    Validation  & 33  & 63  & 1391 & 81  & 829 & 482    & 25&1360&7    & 15&1377&0 \\ 
    Test (Reports)& 32  & 66  & 1430  & 92&646&692   & 22&1408&0    & 11&1419&0 \\
    Test (Expert) & 32  & 66  & 1430  &  374&1056&0    & 46&1384&0     & 31&1399&0  \\
    \end{tabular*}
    \centering
    \captionsetup{type=table}
    \caption{The dataset used to train SCT to detect metastases, fractures and compression. 
    The first-three rows indicate the number 
    of labels extracted from free-text reports for the training, validation and test splits. 
    The bottom row indicates the same test dataset with
    each vertebra independently annotated by an expert. Vertebra are labelled as 
    positive ($+$), negative ($-$), or unknown ($?$). Note the expert annotator was required
    to make their best guess in cases of uncertainty.}
    \label{tab:cancer-dataset-stats}
    \end{footnotesize}
    \end{minipage}
    \end{figure}

A particular challenge of using annotations derived from reports
for training is dealing with ambiguities in the text. Firstly,
reports are often inconclusive; e.g. ``The nature of this lesion is indeterminate''
or  ``Metastases are a possibility but further investigation is required''.
In these cases we label vertebrae as `unknown' for a specific condition and do
not apply a loss to the model's corresponding predictions. Note if one vertebra is marked as positive in a spine 
and the other levels are ambiguous, then the others are marked as `unknown'. 
This accounts for cases where a specific metastasis/fracture 
is being commented on due to a change from a previous scan 
but other unchanged metastases/fractures are not fully described.
Cord compression, on the other hand, is marked as negative unless 
explicitly stated otherwise,  as it is a severe condition 
which is highly unlikely to go unmentioned in a report. Secondly, reports often  state that ``metastatic disease is widespread'' throughout the
spine but do not indicate at which specific levels this disease occurs. 
To derive annotations from these ambiguous reports,
we introduce an additional, global label for each study, indicating if the given condition is
present in the scan at any vertebral level or not. We then aggregate each vertebra-level prediction
to produce a spinal-column level prediction of whether the condition is present.
This setting is an example of multiple-instance learning (MIL), whereby instead of each training instance being individually labelled, sets of 
instances (known as `bags') are assigned a single label, indicating if at least one
instance in the bag is positive. The challenge is then to determine which instances in 
the bag are positives. In this case each bag represents an entire spinal column and each 
instance is a single vertebra. We can then train by a hybrid single-instance and 
multiple-instance learning approach; vertebra-level annotations are used for supervision where given
by the report in addition to the scan-level annotations given in each case. 
A breakdown of the labels extracted by this method is given in Table~\ref{tab:cancer-dataset-stats}.
Examples of vertebral bodies from each class are shown in Figure~\ref{fig:example-cancer-annotations}.
Further explanation and example annotations are given in the appendix.
To ensure our model produces results similar to those given by bespoke
annotations, our test dataset is labelled by an expert spinal surgeon.

\section{Radiological Grading Of Spinal Degenerative Changes}
\label{sec:radiological-grading}

To validate SCT on an existing vertebra analysis problem, we also measure its performance 
at grading common degenerative changes around the intervertebral discs. 
Specifically, we grade the following 
common conditions:  Pfirrmann Grading, Disc Narrowing, Endplate Defects (Upper/Lower), 
Marrow Changes (Upper/Lower), Spondylolisthesis and Central Canal Stenosis. 
We do this on Genodisc, a dataset of clinical lumbar MRIs of 2295 patients from 6 different clinical imaging sites. 
This problem setting closely follows that discussed in~\cite{Jamaludin17b} and~\cite{Windsor22}, 
which give the existing state-of-the-art methods for many of these grading tasks. Each vertebral disc from L5/S1 to T12/L1 in each scan is graded
by an expert radiologist for the aforementioned gradings.
In this setting, SCT takes as input volumes surrounding intervertebral
discs (IVD) and outputs predictions for each grading. The study protocol for scans vary, though the vast majority of
scans have at least a T1-weighted and T2-weighted sagittal scan. In this work we
only consider sagittally sliced scans. However,  there is no reason why SCT 
could not also operate on axial scans in conjunction. A more complete breakdown 
of the Genodisc dataset is given in~\cite{Jamaludin17a}. 
We compare to SpineNet V1~\cite{Jamaludin17b} and SpineNet V2~\cite{Windsor22}, existing
models trained on the same dataset.

\section{Experimental Results}
This section describes experiments to  evaluate the performance of SCT at the 
tasks of: (a) detecting spinal cancer; (b) grading common degenerative changes.
\paragraph{Preprocessing, Implementation \& Training Details:} For both tasks, vertebrae are detected  and labelled
in the scans using the automated method described in~\cite{Windsor22}. This is then used to extract  
volumes surrounding VBs (for the cancer task) and IVDs (for the 
grading task). Volumes are resampled to $S\times112\times112$ and 
$S\times112\times224$ respectively, where $S$ is the number of sagittal slices.
Additional training details are given in the appendix.

\paragraph{Baselines:} For each task, we compare SCT to baseline models operating on 
a single vertebra/IVD at a time.  
For the cancer task we compare to the visual encoder shown in Figure~\ref{fig:vert-transformer-encoder} operating alone
on a single sequence (since this cannot be trivially extended to multiple sequences).
For the radiological grading task we compare with SpineNet V1~\cite{Jamaludin17b}, SpineNet V2~\cite{Windsor22} and the visual encoder alone.
We also train SCT on T1 and T2 scans independently.

\begin{figure}[h!]
    \centering
    \begin{minipage}[h]{0.45\linewidth}
        \includegraphics[width=\linewidth]{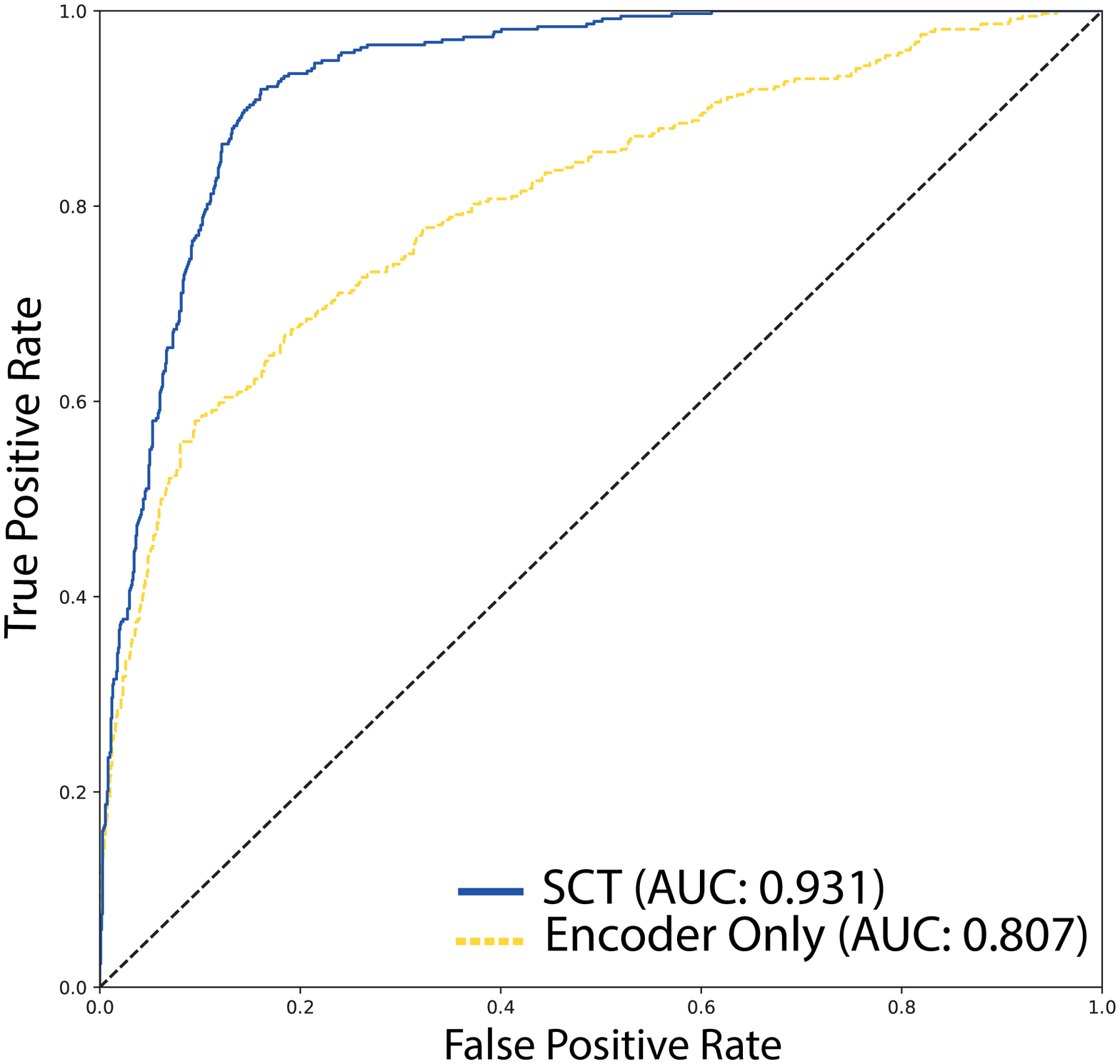}
        \caption{ROC curves for the metastases task
        for SCT and the encoder baseline.
        Curves for the other tasks are given in the appendix.}  
        \label{fig:spine-cancer-rocs}
    \end{minipage}
    \begin{minipage}[h]{0.45\linewidth}
        \centering
        \captionsetup{type=table}
        \caption{AUC scores for the three tasks with various models. We compare to the
            report-extracted annotations and also expert annotations of each image.}
            Expert-Labelled Test Set Annotations
        \begin{tabular}{|c| c |c |c |}
            \hline
             & \multicolumn{3}{c|}{ROC AUC} \\
            \hline
             & Mets & Frac. & Cmprs. \\
             \hline
             \
             Baseline & 0.80 & 0.975 & \textbf{0.930} \\
             SCT & \textbf{0.931} & \textbf{0.980} & 0.868  \\
             \hline
        \end{tabular}
        \\
        \ \\
            Report-Extracted Test Set Annotations
        \begin{tabular}{|c| c |c |c |}
            \hline
             & \multicolumn{3}{c|}{ROC AUC} \\
            \hline
             & Mets & Frac. & Cmprs. \\
             \hline
             \
             Baseline & 0.934 & 0.902 & \textbf{0.955} \\
             SCT & \textbf{0.944} & \textbf{0.901} & 0.918  \\
             \hline
        \end{tabular}
        \label{tab:spine-cancer-results}
    \end{minipage}
\end{figure}

\paragraph{Results:} Table~\ref{tab:spine-cancer-results} shows the performance of all models at the cancer task.
The SCT model performs well across all tasks. In particular, it 
performs much better at detecting subtle metastases
(AUC: \textbf{0.80$\rightarrow$0.931} for the expert labels). This effect can be 
seen clearly in Figure~\ref{fig:spine-cancer-rocs}. This makes sense
as patients will often present with multiple metastases and thus obvious metastases 
in one area of the spine will inform predictions on marginal cases elsewhere in the spine.
We note slightly depreciated performance at the compression task. We believe this
is due to overfitting as there are relatively few compression cases in the training 
set.

Table~\ref{tab:genodisc-results} shows the performance of all
models at the radiological grading tasks, and compares to the state-of-the-art SpineNet model~\cite{Jamaludin17a}. 
SCT outperforms  SpineNet V1 \& V2  on the same dataset for all
tasks except central canal stenosis where the difference in performance is minimal . 
In total average performance increases from \textbf{85.9\%}$\rightarrow$\textbf{87.4}\%.
As expected, the multiple sequence model exceeds  the performance of single sequence models 
in most tasks. The largest improvements can be seen in the endplate defect task 
 (\textbf{82.9/87.8\%} $\rightarrow$\textbf{87.2/90.7\%}),
disc narrowing (\textbf{76.1\%}$\rightarrow$\textbf{77.4\%}) 
and Pfirrmann grading (\textbf{71.0\%} $\rightarrow$\textbf{73.0\%}).
The Pfirrmann grading score of the T1-only model is far worse than the models
with T2 sequences (64.5\%).
This matches expectations as one of the criteria of Pfirrmann grading is the intensity of the intervertebral discs in T2 sequences~\cite{pfirrmann_magnetic_2001}. Overall, the improved
performance is a clear indication of the benefit of considering context from 
multiple sequences and IVDs together. 
\begin{table}[h!]
     \centering
     \begin{footnotesize}

      \begin{tabular*}{\textwidth}{@{\extracolsep{\fill}}lcccc}
         \hline
                                                                               & Pfirrmann & Disc Narrowing         & C.C.S. & Spondylolisthesis \\
     Model \hfill \# Classes:                                                                & 5         & 4              & 2               & 2      \\
     \hline
     SpineNet V1 (T2)~\cite{Jamaludin17}                                       & 71.0 & 76.1           & \textbf{95.8}            & 95.4   \\
     SpineNet V2 (T2)~\cite{Windsor22}                                               & 70.9 & 76.3           & $93.2^\dagger$            & 95.0   \\
     Baseline: SCT Encoder (T2)                                                 & 70.8      & 74.6           & 94.3            & 96.2   \\
     SCT (T1)                                                      & 64.5      & 73.9           & 93.6            & \textbf{96.6}   \\
     SCT (T2)                                                      & 71.7      & 76.9  & 93.6            & 96.2   \\
     SCT (T1,T2)                                                   & \textbf{73.0}      & \textbf{77.4}           & 94.9            & 95.5   \\
     \hline
     
     \end{tabular*}

      \begin{tabular*}{\textwidth}{@{\extracolsep{\fill}}lcccccc}
         \hline
                                       & \multicolumn{2}{c}{Endplate Defect } & \multicolumn{2}{c}{Marrow Change} & \textbf{Average}\\
                                       & Upper & Lower                           & Upper & Lower                &\\
     Model \hfill \# Classes:           & 2 & 2             & 2 & 2             & \\
     \hline
     SpineNet V1 (T2)~\cite{Jamaludin17} & 82.9       & 87.8           & 89.2                 & 88.4       &  85.8 \\
     SpineNet V2 (T2)~\cite{Windsor22}      & 84.9          & 89.8           & 88.9                 & 88.2       &  85.9 \\
     Baseline: SCT Encoder (T2)        & 87.2          & 88.2           & 90.0                 & 89.2                &  86.3 \\
     SCT (T1)                          & 84.7          & 87.0           & 88.4                 & 88.7                &  84.7 \\
     SCT (T2)                          & 85.3          & 89.3           & \textbf{91.0}        & \textbf{90.2}       &  86.8 \\
     SCT (T1,T2)                       & \textbf{87.2} & \textbf{90.7}  & 90.1                 & 89.9                &  \textbf{87.4} \\
     \hline
     \end{tabular*}
 \caption{Results of the IVD grading task. 
 The balanced accuracy for each sub-task
 is shown. CCS represents central canal stenosis. 
 \textsuperscript{$\dagger$}For SpineNetV2, CCS is originally graded with 4 degrees of severity. We combine the severity classes 2-4 (mild, moderete \& severe CCS) into a single class compared to 
 class 1 (no CCS).}
 \label{tab:genodisc-results}
 \end{footnotesize}
 \end{table}
\section{Conclusion}
 In this paper we present SpinalContextTransformer (SCT) for the analysis
of  multiple vertebrae and multiple MRI sequences.
We demonstrate SCT being applied to a new 
task, detecting conditions related to spinal cancer including metastases, vertebral
collapses and metastatic cord compression. 
We also show that the model improves on existing models for radiological grading of
common spinal conditions and can be used flexibly with varying input imaging modalities.
Finally, it is worth noting that the SCT architecture is applicable to the analysis of other modalities (e.g. CT, X-Ray), fields-of-view (e.g. coronal, axial)
and to other repeated anatomical  
sequential structures, such as teeth or ribs.

\paragraph{Acknowledgements and Ethics:} 
 Ethics for the spinal cancer dataset extraction are provided by OSCLMRIC (IRAS project ID: 207857). 
 We are grateful to Dr.\ Sarim Ather, Dr.\ Jill Urban and Prof.\ Jeremy Fairbank
 for insightful conversations on the clinical aspect of this work as well as Prof.\ Ian McCall 
 for annotating the data.
 Finally, we thank to our funders: Cancer Research UK via the EPSRC AIMS CDT and EPSRC Programme Grant
 Visual AI (EP/T025872/1).

\bibliographystyle{splncs04}
\bibliography{shortstrings,vgg_local,other}

\section{Appendix}
\def\checkmark{\tikz\fill[scale=0.4](0,.35) -- (.25,0) -- (1,.7) -- (.25,.15) -- cycle;} 
\definecolor{codegreen}{rgb}{0,0.6,0}
\definecolor{codegray}{rgb}{0.5,0.5,0.5}
\definecolor{codepurple}{rgb}{0.58,0,0.82}
\definecolor{backcolour}{rgb}{0.95,0.95,0.92}

\lstdefinestyle{mystyle}{
    backgroundcolor=\color{backcolour},   
    commentstyle=\color{codegreen},
    keywordstyle=\color{magenta},
    numberstyle=\tiny\color{codegray},
    stringstyle=\color{codepurple},
    basicstyle=\ttfamily\tiny,
    breakatwhitespace=false,         
    breaklines=true,                 
    captionpos=b,                    
    keepspaces=true,                 
    numbers=left,                    
    numbersep=5pt,                  
    showspaces=false,                
    showstringspaces=false,
    showtabs=false,                  
    tabsize=2
}
\subsection{Pytorch-Style Pseudocode for SCT}
\lstset{style=mystyle}
\begin{lstlisting}[language=Python]
class SpinalContextTransformer():
    def initialize(self, n_sequences, n_vertebrae):
        self.visual_encoder = VisualEncoderModel() # extracts features from each vertebra
        self.transformer    = TransfomerEncoderLayers(n_layers=2) # transforms features using context
        self.E              = 128 # embedding/token size for transformer
        # embedders convert one-hot encoded vert level/positions and type MR sequences
        # into something that can be summed with the features extracted from each vertebra
        self.positional_embedder = LinearLayer(in_dim=n_sequences, out_dim=self.E)
        self.sequence_embedder   = LinearLayer(in_dim=n_vertebrae, out_dim=self.E)
        
        self.sequence_attention = LinearLayer(in_dim=self.E, out_dim=self.E)
        # output classification layers
        self.met_classifier         = LinearLayer(in_dim=self.E, out_dim=1)
        self.fracture_classifier    = LinearLayer(in_dim=self.E, out_dim=1)
        self.compression_classifier = LinearLayer(in_dim=self.E, out_dim=1)

    def forward(self, vert_vols, one_hot_position_encoding, one_hot_sequence_encoding):
        B, N, C, S, H, W = vert_vols.shape
        B, N, C, n_sequences = one_hot_sequence_encoding.shape
        B, N, C, n_vertebrae = one_hot_position_encoding.shape

        # B=Batch size, N=Num Vertebrae, C=Num Channels/MR Sequences, 
        # S=Number of Slices, H=Height, W=Width
        vert_vols = vert_vols.flatten(start_dim=0, end_dim=2) 
        # [B,N,C,S,H,W] -> [B*N*C,S,H,W]
        vert_features = self.visual_encoder(vert_vols)
        # vert_features has dimension [B*N*C,E] where E is the embedding size (E=128 here)
        vert_features        = vert_features.reshape(B,N,C,S,self.E)
        positional_embedding = self.positional_embedder(one_hot_position_encoding)
        sequence_embedding   = self.sequence_embedder(one_hot_sequence_encoding)
        # add in positional and sequence embedding
        input_tokens  = vert_features + positional_embedding + sequence_embedding
        # used transformer to consider context between all vertebrae and sequences
        output_tokens = self.transformer(input_tokens)
        output_tokens = output_tokens.reshape(B,N,C,self.E)
        # now pool along MR sequence dimension (C)
        sequence_attention_scores = self.sequence_attention(output_tokens)
        # softmax across sequence dimension
        sequence_attention_weights = SoftMax(sequence_attention_scores,dim=2) 
        pooled_output_tokens = (output_tokens*sequence_attention_weights).sum(dim=2)
        # pooled_output_tokens has dimension [B,N,E]. 
        # Now apply lineaer classifiers to get output of shape [B,N,1] for each task
        met_score      = self.met_classifier(pooled_output_tokens)
        fracture_score = self.fracture_classifier(pooled_output_tokens)
        mscc_score     = self.compression_classifier(pooled_output_tokens)
        return met_score, fracture_score, mscc_score
\end{lstlisting}

\subsection{Additional Model Training Details}
 All models are trained until 10 successive validation
epochs do not result in a decreased loss. An Adam optimizer is used with a learning rate
of $10^{-4}$ and $\mathbf{\beta}=(0.9,0.999)$. A batch size of 20 
is used for the lumbar-only degenerative changes task and 6 for the whole-spine cancer task
(each sample consists of each VB in each MR sequence the scan). 
In total training for both tasks consumes around 40GB of GPU memory and 
takes approximately 6 hours using 2 Tesla P40s. 
The classification linear layers are then finetuned for each
task for 10 epochs with the encoder and transformer weights frozen. Standard augmentations
for each vertebra are used including rotation ($\pm15\degree$), translation ($\pm32$px), 
scaling ($\pm10\%$) and intensity augmentation ($\pm10\%$). For the multi-sequence
models trained on T1 and T2 sequences, a single sequence or both sequences are dropped during training 
with probabilities of 0.4 and 0.1 respectively, leaving the
model make predictions for a volume it has not seen based on context alone. 50\% dropout is used for the transformer layers.
Pytorch 1.10 was used to implement all models.

\subsection{Slicewise Predictions}

\begin{figure}[H]
    \centering
    \includegraphics[width=.9\linewidth]{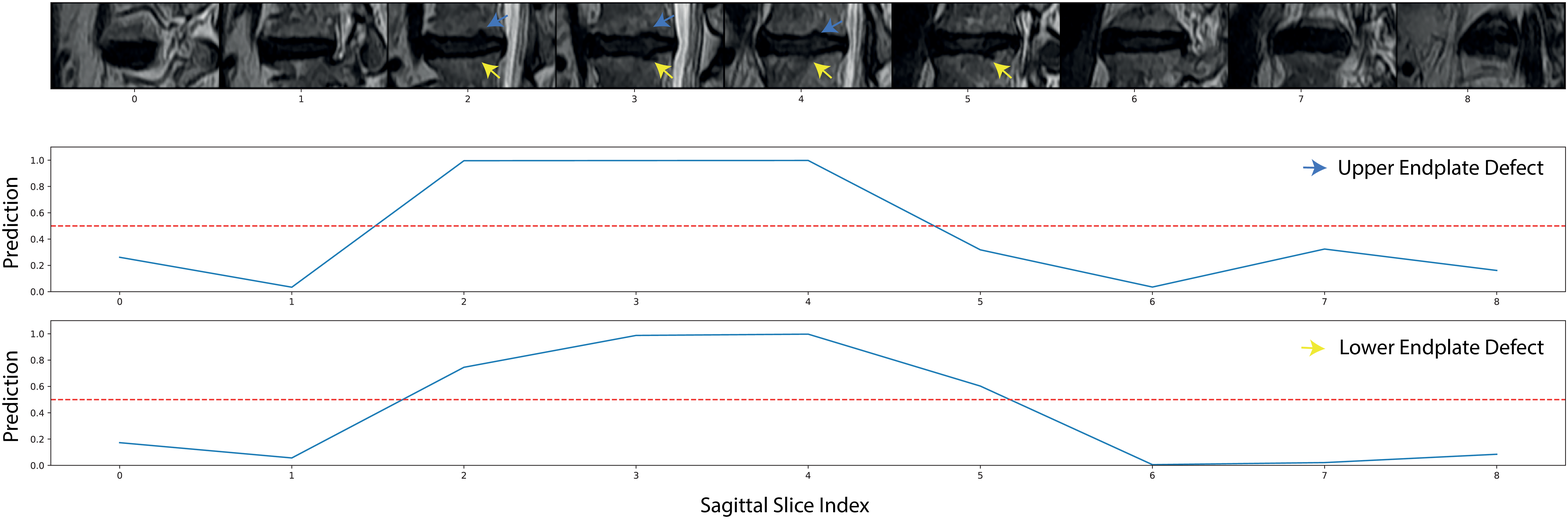}

    \caption{Slicewise outputs for SCT operating on an IVD (top row) with
    upper (blue arrows) and lower endplate defects (yellow arrows). 
    The second and third rows show the predictions for upper and lower endplate defects 
    respectively when each sagittal slice of the scan is fed into the model independently.
    SCT is highly accurate at detecting pathology even
    at the single slice level, detecting the upper endplate defects in slices 2,3 \& 4 and
    the lower endplate is slices 2,3,4 \& 5. This can considered a form of attribution.
    The dotted red line shows $p(\textrm{endplate defect})=50\%$ for both slicewise predictions.}
\end{figure}

\begin{figure}[htb!]
    \centering
    \includegraphics[width=.9\linewidth]{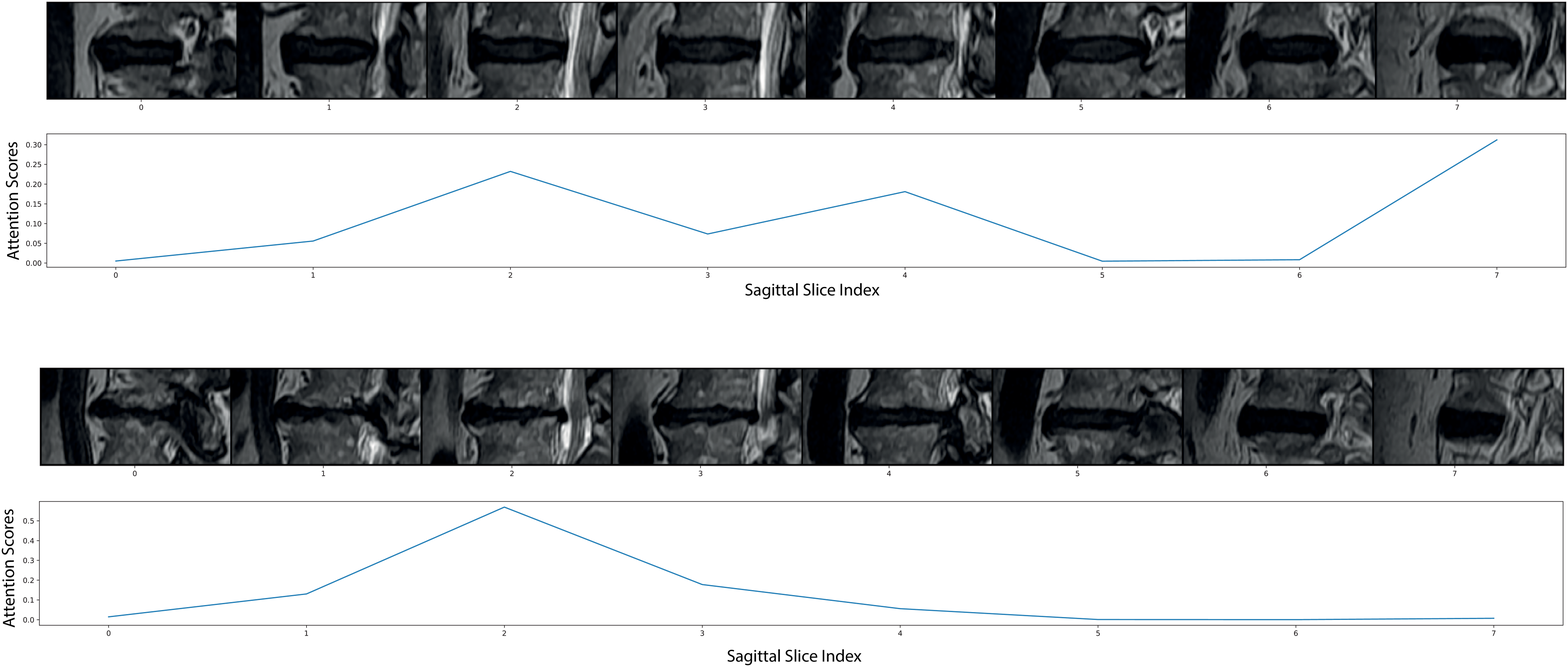}
    \caption{Slicewise attention values for a healthy (top) and an unhealthy (bottom) IVD.
    In the healthy IVD, attention is spread across all mid-sagital slices whereas in the unhealthy 
    vertebra, the attention is concentrated on slices which clearly show pathology (endplate defects,
    narrowing and slight stenosis in slices 1,2 \& 3).}
\end{figure}

\subsection{Example Annotation}

\begin{figure}[H]
    \centering
    \includegraphics[width=.8\linewidth]{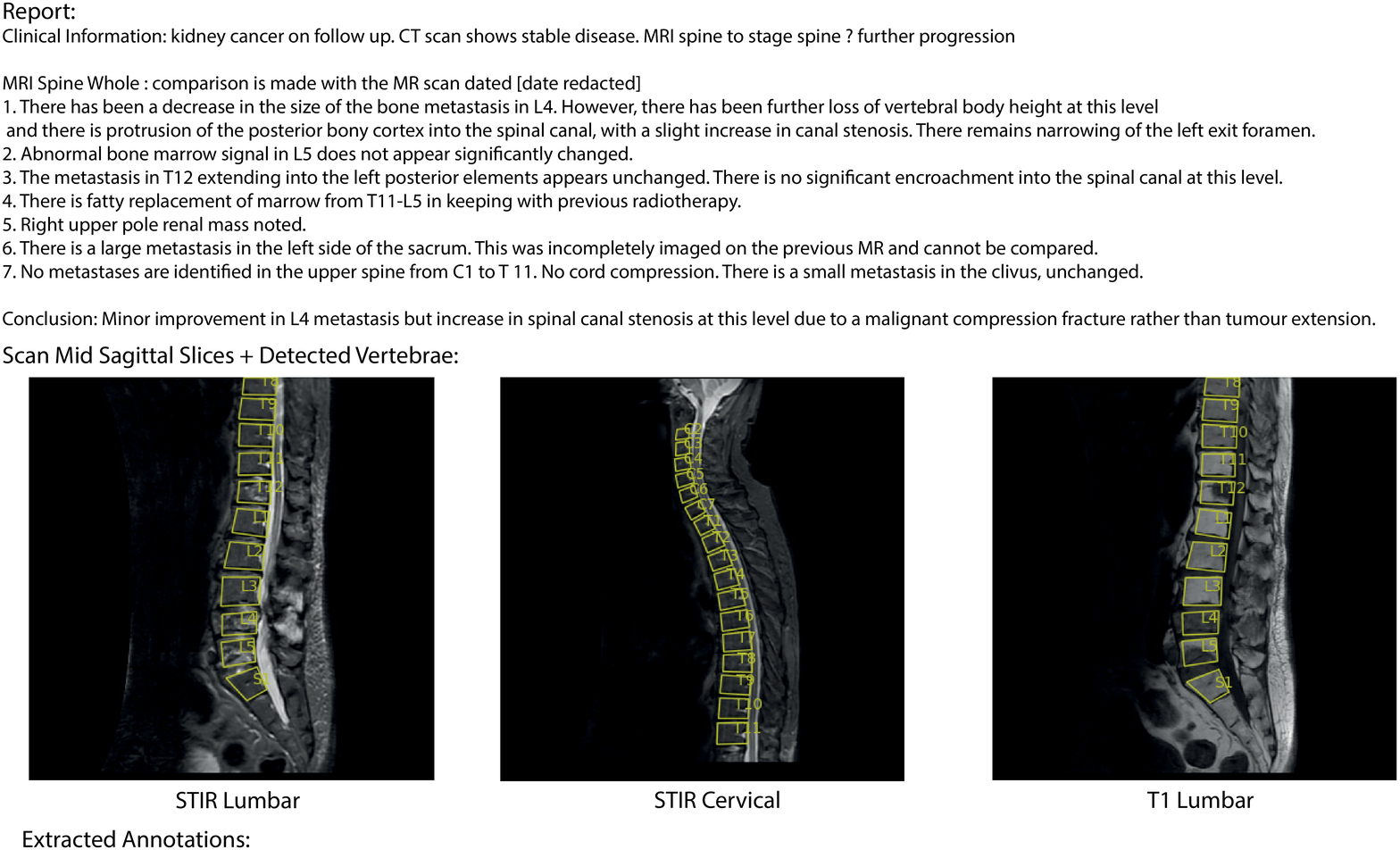}
    \begin{footnotesize}
     \begin{tabular}{@{\extracolsep{\fill}}|c|c|c|c|c|c|c|c|c|c|c|c|c|c|c|c|c|c|c|c|c|c|c|c|c|c|}
        
     \hline
     Condition          & Widespread? & S1 & L5 & L4         & L3 & L2 & L1 & T12       & T11 & T10 & T9 \\
     \hline

     Metastasis?        & \checkmark  &    &    & \checkmark &    &    &    & \checkmark &     &     &   \\
     Fracture?          &             &    &    & \checkmark &    &    &    &            &     &     &   \\
     Compression?       &             &    &    & \checkmark &    &    &    &            &     &     &   \\
     \hline
     \end{tabular}
     \\
     \begin{tabular}{@{\extracolsep{\fill}}|c|c|c|c|c|c|c|c|c|c|c|c|c|c|c|c|c|c|c|c|c|c|c|c|c|c|}
        
     \hline
     Condition          & T8 & T7  & T6 & T5 & T4 & T3 & T2 & T1 & C7 & C6 & C5 & C4 & C3 & C2 \\
     \hline

     Metastasis?        &     &    &    &    &    &    &    &    &    &    &    &    &    &   \\
     Fracture?          &     &    &    &    &    &    &    &    &    &    &    &    &    &   \\
     Compression?       &     &    &    &    &    &    &    &    &    &    &    &    &    &   \\
     \hline
     \end{tabular}
    \end{footnotesize}
    \caption{An example of annotations extracted from a free-text report. There are metastases
    at T12 and L4, as well as cord compression and fractures (``malignant compression fracture'') at L4. 
    The report also suggests there maybe other metastases, e.g. in L5, however these are not explicitly stated
    to be present.
    For this reason the `widespread' column is ticked for metastases. As a results 
    all vertebrae other than T12 and L4 (i.e. levels where metastases are 
    explicitly mentioned) are treated as `unknown'.}
\end{figure}

\subsection{ROC Curves for All Cancer Tasks}

\begin{figure}[H]
    \begin{subfigure}{0.32\textwidth}
        \includegraphics[width=\linewidth]{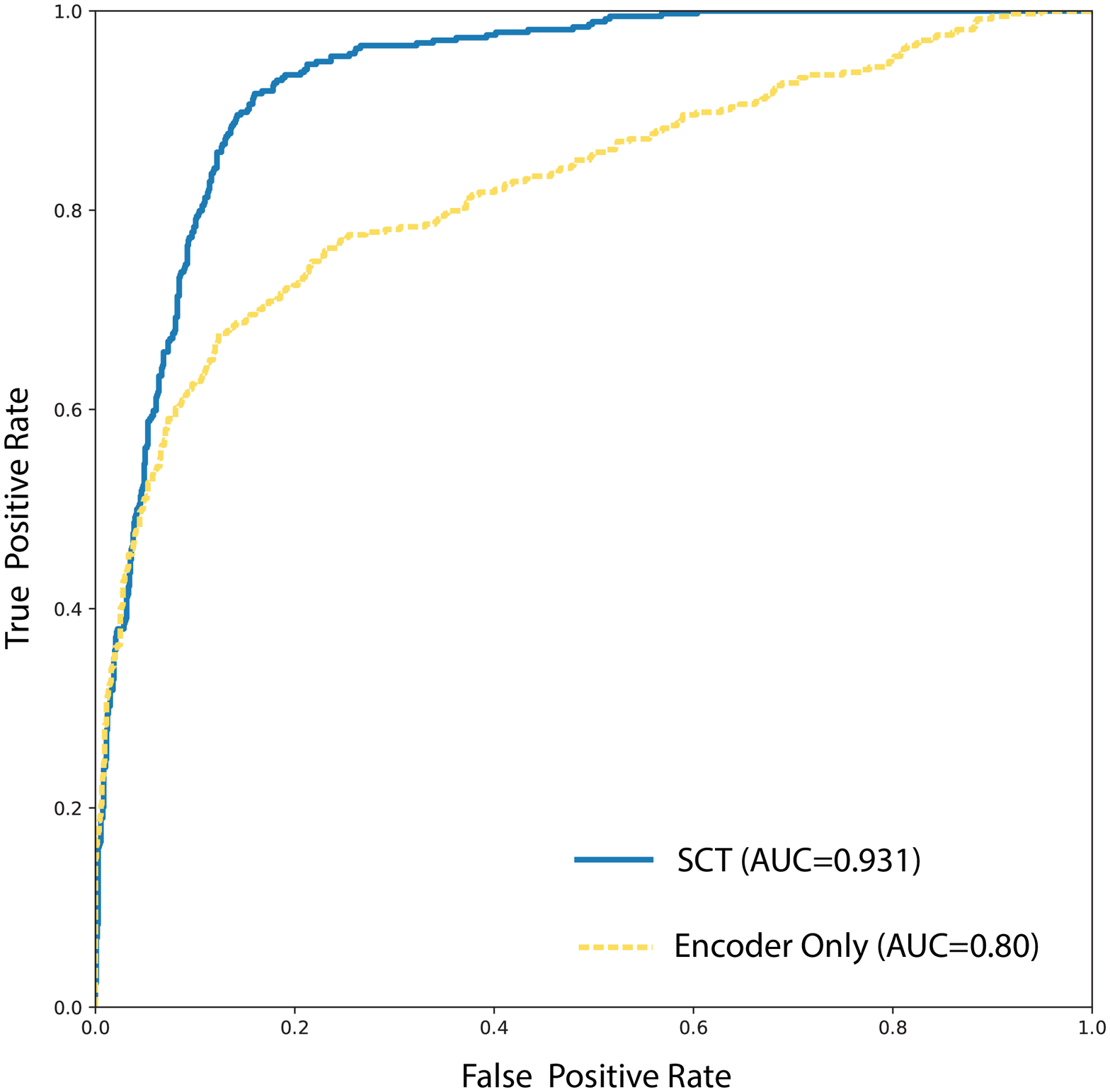}
        \caption{Metastasis}
    \end{subfigure}
    \begin{subfigure}{0.32\textwidth}
        \includegraphics[width=\linewidth]{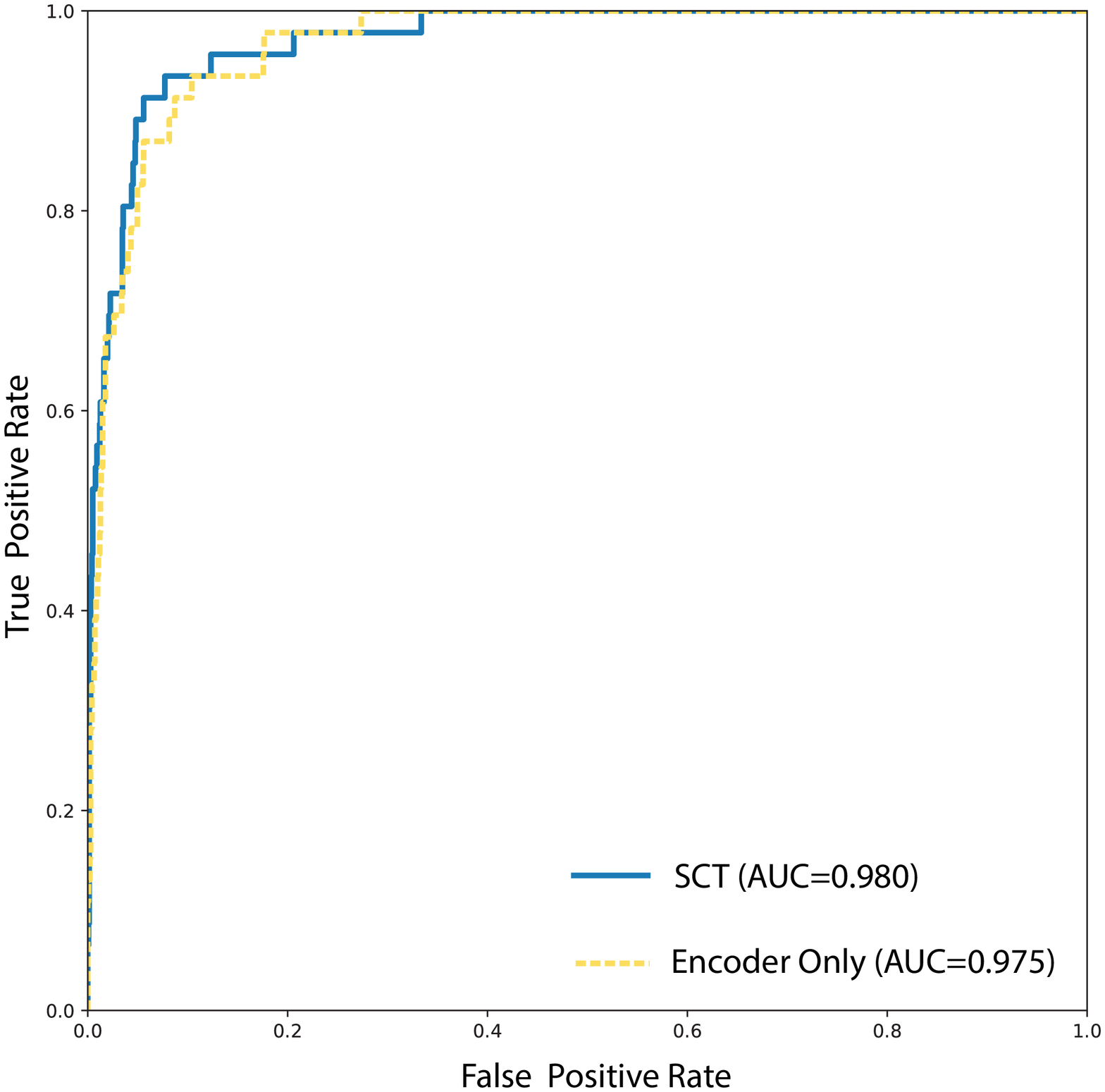}
        \caption{Vertebral Fracture}
    \end{subfigure}
    \begin{subfigure}{0.32\textwidth}
        \includegraphics[width=\linewidth]{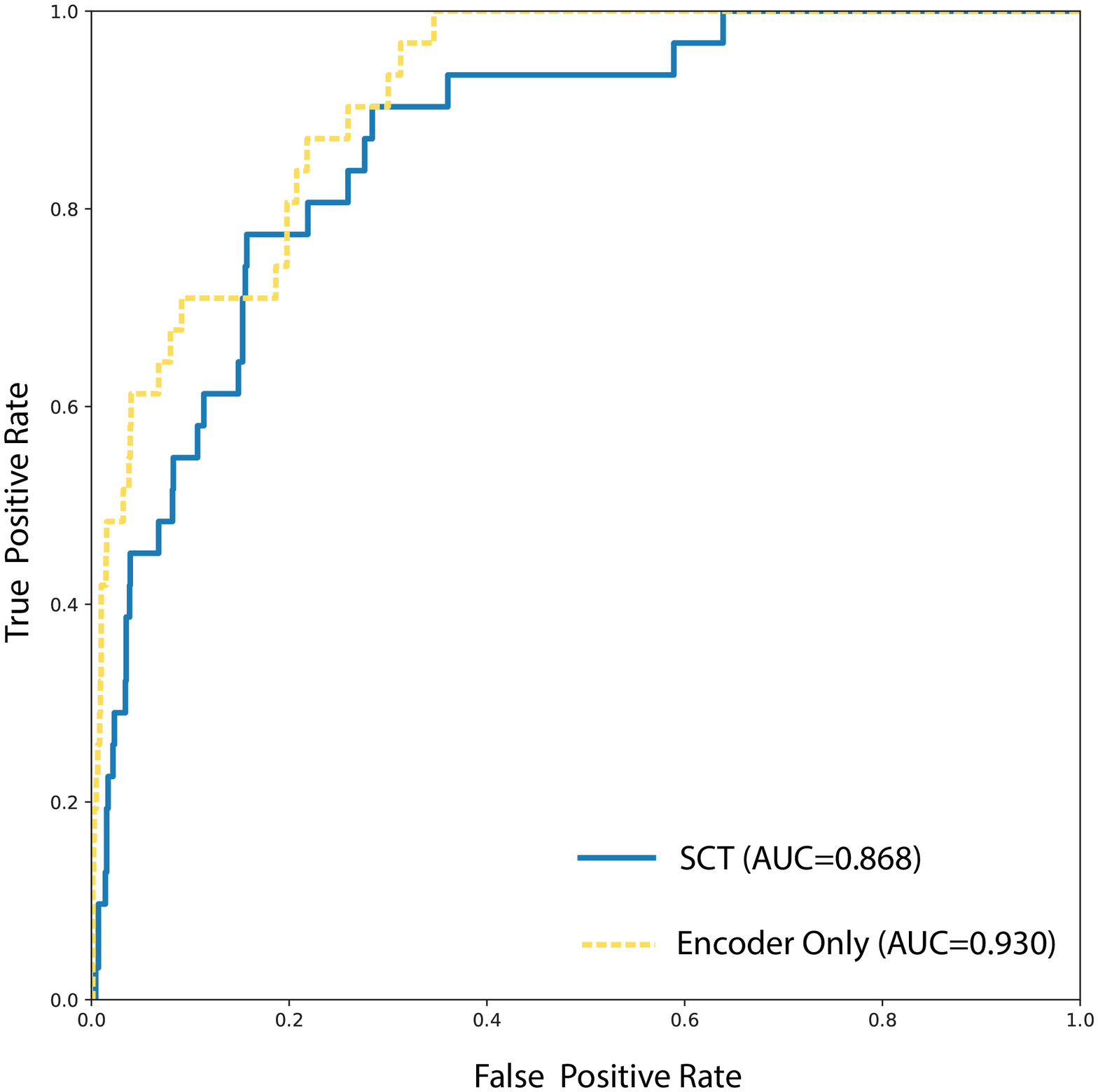}
        \caption{Cord Compression}
    \end{subfigure}
    \caption{ROC Curves for the individual spine cancer tasks. These
    curves show the agreement of the model's predictions with the annotations
    produced by an expert spinal surgeon.}
\end{figure}

\end{document}